\documentclass[12pt]{article}
\usepackage{}
\usepackage{amscd}
\usepackage{bbm}
\usepackage{mathrsfs}
\usepackage{amssymb}
\usepackage{graphics}
\usepackage{graphicx}
\usepackage{amsfonts}
\usepackage{amsmath}
\allowdisplaybreaks
\usepackage[numbers,sort&compress]{natbib}
\begin{document}
\date{}
\title{Bright-dark vector soliton solutions for the coupled complex short pulse equations in nonlinear optics}
\author{Bo-Ling Guo, Yu-Feng Wang\thanks{Corresponding
author, with e-mail address as
 yufeng$_{-}$0617@126.com} \\
\\
{\em Institute of Applied Physics and Computational Mathematics,}\\{\em Beijing 100088, China}
}
\maketitle \vspace{-0.8cm}
\begin{abstract}
Under investigation in this paper are the coupled complex short pulse equations, which describe the propagation of ultra-short optical pulses in cubic nonlinear media. Through the Hirota method, bright-dark one- and two-soliton solutions are obtained. Interactions between two bright or two dark solitons are verified to be elastic through the asymptotic analysis. With different parameter conditions of the vector bright-dark two solitons, the oblique interactions, bound states of solitons and parallel solitons are analyzed.

\end{abstract}

\newpage

\noindent\textbf{1. Introduction}\\

The propagations of picosecond optical pulses in the single-mode nonlinear media are usually described by the cubic nonlinear Schr\"{o}dinger~(NLS) equation~\cite{APL1973,PRL1980,fiber-book-1995}, but to the ultra-fast signal transmissions, the cubic NLS equation is no longer valid~\cite{fiber-book-1995}. On the one hand, people study the subpicosecond and femtosecond pulses by incorporating higher terms in the cubic NLS equation~\cite{IEEE1987,OL1990,PRL2000}, including the higher-order dispersion, self-steepening and Raman scattering. On the other hand, people try to derive some new equations besides the NLS equation to describe the ultra-short pulses under certain assumptions~\cite{PLA2000,PRL1997,JOSAB-1991,JOSAB-1996,PRE2014,sp}.

The short pulse equation which is derived under the assumption that the pulse center is far from the nearest resonance frequency of the material's susceptibility reads as~\cite{sp}
\begin{eqnarray}\label{sp}
u_{XT}=u+\frac{1}{6}(u^{3})_{XX},
\end{eqnarray}
which describes the propagation of ultra-short optical pulses in cubic nonlinear media, where the real function $u$ of $T$ and $X$ stands for the magnitude of the electric field~\cite{sp}, the subscripts mean the partial differentiations. Eq.~(\ref{sp}) is proved to be completely integrable~\cite{sp-integrable-JPSJ-2005,Hamilton-PLA-2006}, which owns the Lax pair in the Wadati-Konno-Ichikawa type. Solitary solutions for Eq.~(\ref{sp}) have been derived~\cite{JPA2006}. Periodic solutions, multi-loop soliton and multi-breather solutions for Eq.~(\ref{sp}) have been constructed by a hodograph transformation which converts Eq.~(\ref{sp}) into the sine-Gordon equation~\cite{JMP2008,JPSJ-2007}. The integrable discretization and the geometric interpretation of Eq.~(\ref{sp}) have been given in Refs.~\cite{Feng-JPA-2010,Feng-JPA-2011}.

Advantages have been verified during the process of dealing with the NLS equation that the complex function can describe the propagation of optical pulse along the optical fibers more properly than the real one~\cite{optical-waves-1983}. The complex function contains the amplitude and phase, which are two characters of an optical pulse envelop~\cite{optical-waves-1983}. In addition, the complex functions pave the ways to study the soliton interaction~\cite{optical-waves-1983}. The complex form of Eq.~(\ref{sp}) reads as~\cite{complex-sp-Physica-D-2015}
\begin{eqnarray}\label{complex-sp}
q_{xt}+q+\frac{1}{2}(|q|^{2}q_{x})_{x}=0,
\end{eqnarray}
where $q$ is the complex function of $x$ and $t$, which represents the magnitude of the electric field. The derivation of Eq.~(\ref{complex-sp}) from the Maxwell equation was pointed in Ref.~\cite{complex-sp-Physica-D-2015}. Integrability and multi-bright-soliton solutions for Eq.~(\ref{complex-sp}) have been given~\cite{complex-sp-Physica-D-2015}. The link between the complex coupled dispersionless equation and Eq.~(\ref{complex-sp}) has been carried out~\cite{Studies-2015}.

Contrast to the single-component equation, the multi-component ones own the inter-component and inner-component nonlinearity terms~\cite{fiber-book-1995,CNLS-PRL-1997}. Owing to the different polarization of each component may has, vector soliton solutions for the multi-component equations have been displayed in the bright-bright, bright-dark and dark-dark forms~\cite{bright-bright-1,bright-bright-2,dark-dark,bright-dark-1,bright-dark2}. Vector solitons have shown to be applicable in the optical switch and logic gate~\cite{fiber-book-1995,OL1993}, since multiple kinds of interactions between the bright vector solitons have been experimentally and theoretically performed, including the shape-changing and shape-preserving interactions~\cite{bright-bright-1,bright-bright-2}.

To describe the propagation of optical pulses in birefringence fibers, Ref.~\cite{complex-sp-Physica-D-2015} derived the coupled complex short pulse equations~\cite{complex-sp-Physica-D-2015}
\begin{subequations}\label{coupled-complex-sp}
\begin{align}
&q_{1,xt}+q_{1}+\frac{1}{2}\left[(|q_{1}|^{2}+|q_{2}|^{2})q_{1,x}\right]_{x}=0,\label{coupled-complex-sp-a}\\
&q_{2,xt}+q_{2}+\frac{1}{2}\left[(|q_{1}|^{2}+|q_{2}|^{2})q_{2,x}\right]_{x}=0,\label{coupled-complex-sp-b}
\end{align}
\end{subequations}
where $q_{1}$ and $q_{2}$ are the complex functions of $x$ and $t$, which indicate the magnitudes of the electric fields. The Lax pair, conservation laws and bright soliton solutions in pfaffians have been displayed~\cite{complex-sp-Physica-D-2015}. The multi-bright-soliton, multi-breather and higher-order rogue wave solutions for Eqs.~(\ref{coupled-complex-sp}) have been obtained through Darboux transformation~\cite{Ling-2015}.

However, to our knowledge, the bright-dark vector soliton solutions and their interactions for Eqs.~(\ref{coupled-complex-sp}) have not been put forward. Motivated by the above, in this paper, we will study the bright-dark vector soliton solutions and their interactions for Eqs.~(\ref{coupled-complex-sp}) via the Hirota method. In Section~2, the bright-dark one- and two-soliton solutions for Eqs.~(\ref{coupled-complex-sp}) will be obtained based on the bilinear forms under scale transformations. Asymptotic analysis will be performed on the two-soliton solutions in Section~3. Oblique interaction, bound states of solitons and parallel solitons will be analyzed in Section~4. Section~5 will be our conclusions. \\

\noindent\textbf{2. Soliton solutions for Eqs.~(\ref{coupled-complex-sp})}\\

The Hirota method is a directive and effective way to deal with the solutions for the nonlinear evolution equations~(NLEEs)~\cite{Hirota}. Soliton solutions will be obtained through the truncated parameter expansion at different levels based on the bilinear forms, which can be obtained through certain transformations~\cite{Hirota}.

With the dependent variable transformations
\begin{eqnarray}\label{rational-transformation}
q_{1}=\frac{g_{1}}{f}\,e^{i(y-s)},\hspace{1cm}q_{2}=\frac{g_{2}}{f}\,e^{i(y-s)},
\end{eqnarray}
and the scale transformations
\begin{eqnarray}\label{scale-transformation}
x=\lambda\,{y}+2\lambda\,{s}-2(\ln{f})_{s},\hspace{1cm}t=-s,
\end{eqnarray}
the bilinear forms for Eqs.~(\ref{coupled-complex-sp}) will be written as
\begin{subequations}\label{bilinear-forms}
\begin{align}
&(D_{y}D_{s}+iD_{s}-iD_{y}-\lambda-1)g_{1}\cdot\,f=0,\\
&(D_{y}D_{s}+iD_{s}-iD_{y}-\lambda-1)g_{2}\cdot\,f=0,\\
&(D_{s}^{2}-2\lambda)f\cdot\,f=\frac{1}{2}(|g_{1}|^{2}+|g_{2}|^{2}),
\end{align}
\end{subequations}
where $g_{1}$ and $g_{2}$ are both complex functions, $f$ is a real one, $D_{y}$ and $D_{s}$ are the bilinear derivative operators~\cite{Hirota} defined by
\begin{eqnarray}
D_y^{m_1}\,D_s^{m_2}\,a\cdot\,b=\Big(\frac{\partial}{\partial\,y}-\frac{\partial}{\partial\,y'}\Big)^{m_1}\,
\Big(\frac{\partial}{\partial\,s}-\frac{\partial}{\partial\,s'}\Big)^{m_2}\,
a(y,s)b(y',s')\Big|_{y'=y,s'=s},
\end{eqnarray}
with $y$, $s$, $y'$ and $s'$ being formal variables, $a(y,s)$ and $b(y',s')$ being two functions, while $m_{1}$ and $m_{2}$ are two nonnegative integers. Proof of Bilinear Forms~(\ref{bilinear-forms}) will be shown in Appendix A.

To obtain the bright-dark soliton solutions for Eqs.~(\ref{coupled-complex-sp}), we truncate $g_{1}$, $g_{2}$ and $f$ at the formal expansion parameter $\epsilon$ as follows:
\begin{subequations}\label{expansion}
\begin{align}
&g_{1}=\epsilon\,g_{1}^{(1)}+\epsilon^{3}g_{1}^{(3)}+\epsilon^{5}g_{1}^{(5)}+\cdots,\\
&g_{2}=g_{2}^{(0)}\left(1+\epsilon^{2}g_{2}^{(2)}+\epsilon^{4}g_{2}^{(4)}+\epsilon^{6}g_{2}^{(6)}+\cdots\right),\\
&f=1+\epsilon^{2}f_{2}+\epsilon^{4}f_{4}+\epsilon^{6}f_{6}+\cdots,
\end{align}
\end{subequations}
$g_{1}^{(l)}$'s ($l=1,3,5,\ldots$) and $g_{2}^{(\iota)}$'s ($\iota=0,2,4,\ldots$) are the complex functions of $y$ and $s$, while $f_{r}$'s ($r=2,4,6,\cdots$) are the real ones, which will be determined later.

Terminating the expansions as
\begin{eqnarray}
&&g_{1}=\epsilon\,g_{1}^{(1)}=\epsilon\,\delta_{1}\,e^{\theta_{1}},\notag\\
&&g_{2}=g_{2}^{(0)}\left(1+\epsilon^{2}g_{2}^{(2)}\right)=\rho\,e^{i\,b\,s}\left(1+\epsilon^{2}\xi_{1}\eta_{1}e^{\theta_{1}+\theta_{1}^{*}}\right),\notag\\
&&f=1+\epsilon^{2}f_{2}=1+\epsilon^{2}\eta_{1}e^{\theta_{1}+\theta_{1}^{*}},\notag
\end{eqnarray}
we express the one-soliton solutions for Eqs.~(\ref{coupled-complex-sp}) as
\begin{subequations}\label{one-soliton}
\begin{align}
&q_{1}=\frac{1}{2}\delta_{1}e^{i(y-s)}e^{\frac{\theta_{1}-\theta_{1}^{*}-\ln{\eta_{1}}}{2}}
\textrm{sech}\left(\frac{\theta_{1}+\theta_{1}^{*}+\ln{\eta_{1}}}{2}\right),\\
&q_{2}=\rho\,e^{i\,b\,s}e^{i(y-s)}\left[\frac{1+\xi_{1}}{2}-\frac{1-\xi_{1}}{2}
\tanh\left(\frac{\theta_{1}+\theta_{1}^{*}+\ln{\eta_{1}}}{2}\right)\right],\\
&x=\lambda\,{y}+2\lambda\,{s}-(\omega_{1}+\omega_{1}^{*})
\left[1+\tanh\left(\frac{\theta_{1}+\theta_{1}^{*}+\ln{\eta_{1}}}{2}\right)\right],\hspace{0.5cm}t=-s,
\end{align}
\end{subequations}
with $\theta_{1}=k_{1}\,y+\omega_{1}\,s+\nu_{1}$, $k_{1}$, $\nu_{1}$, $\delta_{1}$ and $\rho$ being all the complex constants, while $\lambda$, $\omega_{1}$, $b$, $\xi_{1}$ and $\eta_{1}$ are listed in Appendix B.

Terminating $g_{1}$, $g_{2}$ and $f$ as $g_{1}=\epsilon\,g_{1}^{(1)}+\epsilon^{3}g_{1}^{(3)}$,
$g_{2}=g_{2}^{(0)}\left(1+\epsilon^{2}g_{2}^{(2)}+\epsilon^{4}g_{2}^{(4)}\right)$ and $f=1+\epsilon^{2}f_{2}+\epsilon^{4}f_{4}$, we get the bright-dark two-soliton solutions for Eqs.~(\ref{coupled-complex-sp}),
\begin{eqnarray}\label{two-soliton}
&&q_{1}=\frac{g_{1}^{(1)}+g_{1}^{(3)}}{1+f_{2}+f_{4}}e^{i(y-s)},\hspace{0.5cm}
q_{2}=\frac{g_{2}^{(0)}\left(1+g_{2}^{(2)}+g_{2}^{(4)}\right)}{1+f_{2}+f_{4}}e^{i(y-s)},\\
&&x=\lambda\,{y}+2\lambda\,{s}-2\ln(1+f_{2}+f_{4})_{s},\hspace{0.5cm}t=-s,\notag
\end{eqnarray}
where
\begin{eqnarray}
&&g_{1}^{(1)}=\delta_{1}\,e^{\theta_{1}}+\delta_{2}\,e^{\theta_{2}},\hspace{0.5cm}
g_{1}^{(3)}=\delta_{3}\,e^{\theta_{1}+\theta_{2}+\theta_{1}^{*}}+\delta_{4}\,e^{\theta_{1}+\theta_{2}+\theta_{2}^{*}},\notag\\
&&g_{2}^{(2)}=\xi_{1}\eta_{1}\,e^{\theta_{1}+\theta_{1}^{*}}+\xi_{2}\eta_{2}\,e^{\theta_{1}+\theta_{2}^{*}}+
\xi_{3}\eta_{3}\,e^{\theta_{2}+\theta_{1}^{*}}+\xi_{4}\eta_{4}\,e^{\theta_{2}+\theta_{2}^{*}},\notag\\
&&f_{2}=\eta_{1}\,e^{\theta_{1}+\theta_{1}^{*}}+\eta_{2}\,e^{\theta_{1}+\theta_{2}^{*}}+
\eta_{3}\,e^{\theta_{2}+\theta_{1}^{*}}+\eta_{4}\,e^{\theta_{2}+\theta_{2}^{*}},\notag\\
&&g_{2}^{(4)}=\xi_{5}\eta_{5}\,e^{\theta_{1}+\theta_{1}^{*}+\theta_{2}+\theta_{2}^{*}},\hspace{0.5cm}
f_{4}=\eta_{5}\,e^{\theta_{1}+\theta_{1}^{*}+\theta_{2}+\theta_{2}^{*}},\notag
\end{eqnarray}
with
$\theta_{2}=k_{2}\,y+\omega_{2}\,s+\nu_{2}$, $k_{2}$, $\nu_{2}$ and $\delta_{2}$ being all the complex constants, while the relevant coefficients will be shown in Appendix B.\\

\noindent\textbf{3. Asymptotic analysis}\\

In this part, we will make the asymptotic analysis on Solutions~(\ref{two-soliton}) to investigate the interactions between the bright-dark two solitons.

A. Before the interaction ($s\longrightarrow-\infty$)\\
(a) $\theta_{1}+\theta_{1}^{*}\sim0$, $\theta_{2}+\theta_{2}^{*}\sim-\infty$
\begin{subequations}
\begin{align}
&q_{1}\xrightarrow{}S_{1}^{1-}=\frac{\delta_{1}}{2}\,e^{\frac{\theta_{1}-\theta_{1}^{*}-\ln{\eta_{1}}}{2}}\,e^{i(y-s)}\,
\textrm{sech}\left(\frac{\theta_{1}+\theta_{1}^{*}+\ln{\eta_{1}}}{2}\right),\label{before-interaction-1a}\\
&q_{2}\xrightarrow{}S_{2}^{1-}=\rho\,e^{ibs}\,e^{i(y-s)}\left[\frac{\xi_{1}+1}{2}+\frac{\xi_{1}-1}{2}\,
\textrm{tanh}\left(\frac{\theta_{1}+\theta_{1}^{*}+\ln{\eta_{1}}}{2}\right)\right],\label{before-interaction-1b}
\end{align}
\end{subequations}
(b) $\theta_{2}+\theta_{2}^{*}\sim0$, $\theta_{1}+\theta_{1}^{*}\sim+\infty$
\begin{subequations}
\begin{align}
&q_{1}\xrightarrow{}S_{1}^{2-}=\frac{\delta_{3}}{2\,\eta_{1}}\,
e^{\frac{\theta_{2}-\theta_{2}^{*}-\ln{\frac{\eta_{5}}{\eta_{1}}}}{2}}\,e^{i(y-s)}\,
\textrm{sech}\left(\frac{\theta_{2}+\theta_{2}^{*}+\ln{\frac{\eta_{5}}{\eta_{1}}}}{2}\right),\label{before-interaction-2a}\\
&q_{2}\xrightarrow{}S_{2}^{2-}=\rho\,e^{ibs}\,e^{i(y-s)}\left[\frac{\xi_{5}+\xi_{1}}{2}+\frac{\xi_{5}-\xi_{1}}{2}\,
\textrm{tanh}\left(\frac{\theta_{2}+\theta_{2}^{*}+\ln{\frac{\eta_{5}}{\eta_{1}}}}{2}\right)\right],\label{before-interaction-2b}
\end{align}
\end{subequations}
\hspace*{\parindent}B. After the interaction ($s\longrightarrow+\infty$)\\
(a) $\theta_{1}+\theta_{1}^{*}\sim0$, $\theta_{2}+\theta_{2}^{*}\sim+\infty$
\begin{subequations}
\begin{align}
&q_{1}\xrightarrow{}S_{1}^{1+}=\frac{\delta_{4}}{2\,\eta_{4}}\,
e^{\frac{\theta_{1}-\theta_{1}^{*}-\ln{\frac{\eta_{5}}{\eta_{4}}}}{2}}\,e^{i(y-s)}\,
\textrm{sech}\left(\frac{\theta_{1}+\theta_{1}^{*}+\ln{\frac{\eta_{5}}{\eta_{4}}}}{2}\right),\label{after-interaction-1a}\\
&q_{2}\xrightarrow{}S_{2}^{1+}=\rho\,e^{ibs}\,e^{i(y-s)}\left[\frac{\xi_{5}+\xi_{4}}{2}+\frac{\xi_{5}-\xi_{4}}{2}\,\textrm{tanh}\left(\frac{\theta_{1}+\theta_{1}^{*}+\ln{\frac{\eta_{5}}{\eta_{4}}}}{2}\right)\right],\label{after-interaction-1b}
\end{align}
\end{subequations}
(b) $\theta_{2}+\theta_{2}^{*}\sim0$, $\theta_{1}+\theta_{1}^{*}\sim-\infty$
\begin{subequations}
\begin{align}
&q_{1}\xrightarrow{}S_{1}^{2+}=\frac{\delta_{2}}{2}\,e^{\frac{\theta_{2}-\theta_{2}^{*}-\ln{\eta_{4}}}{2}}\,e^{i(y-s)}\,
\textrm{sech}\left(\frac{\theta_{2}+\theta_{2}^{*}+\ln{\eta_{4}}}{2}\right),\label{after-interaction-2a}\\
&q_{2}\xrightarrow{}S_{2}^{2+}=\rho\,e^{ibs}\,e^{i(y-s)}
\left[\frac{\xi_{4}+1}{2}+\frac{\xi_{4}-1}{2}\,\textrm{tanh}\left(\frac{\theta_{2}+\theta_{2}^{*}+\ln{\eta_{4}}}{2}\right)\right],\label{after-interaction-2b}
\end{align}
\end{subequations}
where $S_{1}^{\tau\mp}$ ($\tau=1,2$) denote the status of the bright soliton $S_{1}$ before ($-$) and after ($+$) the interactions, while $S_{2}^{\tau\mp}$ ($\tau=1,2$) stand for the dark soliton $S_{2}$ before ($-$) and after ($+$) the interactions, respectively. The related parameters are shown in Appendix B.\\

Table~1. Physical quantities of bright soliton $S_1$ and dark soliton $S_2$ before and after the interactions.

\

\hspace{-0.7cm}\begin{tabular}{|c c c|c c c|}
\hline
Soliton & Amplitude/Depth & Velocity & Soliton & Amplitude/Depth & Velocity\\[1mm]
\hline $S_1^{1-}$ &
$\sqrt{\frac{\delta_{1}\delta_{1}^{*}}{4\eta_{1}}}$ &
$\frac{\rho\rho^{*}-8}{4(k_{1}k_{1}^{*}-ik_{1}+ik_{1}^{*}+1)}$ &
$S_1^{1+}$ &
$\sqrt{\frac{\delta_{4}\delta_{4}^{*}}{4\eta_{4}\eta_{5}}}$ &
$\frac{\rho\rho^{*}-8}{4(k_{1}k_{1}^{*}-ik_{1}+ik_{1}^{*}+1)}$\\[3mm]
\hline $S_2^{1-}$ &
$|\rho|\sqrt{\frac{\xi_{1}-1}{2}\frac{\xi_{1}^{*}-1}{2}}$ &
$\frac{\rho\rho^{*}-8}{4(k_{1}k_{1}^{*}-ik_{1}+ik_{1}^{*}+1)}$ &
$S_2^{1+}$
& $|\rho|\sqrt{\frac{\xi_{5}-\xi_{4}}{2}\frac{\xi_{5}^{*}-\xi_{4}^{*}}{2}}$ &
$\frac{\rho\rho^{*}-8}{4(k_{1}k_{1}^{*}-ik_{1}+ik_{1}^{*}+1)}$\\[3mm]
\hline $S_1^{2-}$ &
$\sqrt{\frac{\delta_{3}\delta_{3}^{*}}{4\eta_{1}\eta_{5}}}$ &
$\frac{\rho\rho^{*}-8}{4(k_{2}k_{2}^{*}-ik_{2}+ik_{2}^{*}+1)}$ &
$S_1^{2+}$&
$\sqrt{\frac{\delta_{2}\delta_{2}^{*}}{4\eta_{4}}}$ &
$\frac{\rho\rho^{*}-8}{4(k_{2}k_{2}^{*}-ik_{2}+ik_{2}^{*}+1)}$\\[3mm]
\hline  $S_2^{2-}$ &
$|\rho|\sqrt{\frac{\xi_{5}-\xi_{1}}{2}\frac{\xi_{5}^{*}-\xi_{1}^{*}}{2}}$ &
$\frac{\rho\rho^{*}-8}{4(k_{2}k_{2}^{*}-ik_{2}+ik_{2}^{*}+1)}$ &
$S_2^{2+}$ &
$|\rho|\sqrt{\frac{\xi_{4}-1}{2}\frac{\xi_{4}^{*}-1}{2}}$ &
$\frac{\rho\rho^{*}-8}{4(k_{2}k_{2}^{*}-ik_{2}+ik_{2}^{*}+1)}$\\[3mm]
\hline
\end{tabular}\\\\

Through the parameters listed in Appendix B, we can obtain that
\begin{eqnarray}
&\sqrt{\frac{\delta_{1}\delta_{1}^{*}}{4\eta_{1}}}=\sqrt{\frac{\delta_{4}\delta_{4}^{*}}{4\eta_{4}\eta_{5}}},\hspace{0.5cm}
|\rho|\sqrt{\frac{\xi_{1}-1}{2}\frac{\xi_{1}^{*}-1}{2}}=|\rho|\sqrt{\frac{\xi_{5}-\xi_{4}}{2}\frac{\xi_{5}^{*}-\xi_{4}^{*}}{2}},\notag\\
&\sqrt{\frac{\delta_{3}\delta_{3}^{*}}{4\eta_{1}\eta_{5}}}=\sqrt{\frac{\delta_{2}\delta_{2}^{*}}{4\eta_{4}}},\hspace{0.5cm} |\rho|\sqrt{\frac{\xi_{5}-\xi_{1}}{2}\frac{\xi_{5}^{*}-\xi_{1}^{*}}{2}}=|\rho|\sqrt{\frac{\xi_{4}-1}{2}\frac{\xi_{4}^{*}-1}{2}},\notag
\end{eqnarray}
which mean the soliton amplitudes/depths $|S_{1}^{1-}|=|S_{1}^{1+}|$, $|S_{2}^{1-}|=|S_{2}^{1+}|$, $|S_{1}^{2-}|=|S_{1}^{2+}|$ and $|S_{2}^{2-}|=|S_{2}^{2+}|$ can be verified through calculations. From Table~1, we are easy to find that the velocities for both the bright and dark soltions keep unchanged before and after the interactions. Therefore, the interactions between two bright or two dark solitons are both elastic.\\

\noindent\textbf{4. Soliton interactions}\\

We will analyze the soliton interactions based on Solutions~(\ref{two-soliton}) in the $y$-$s$ plane. Due to the soliton velocities and distance between solitons, kinds of interactions will be analyzed: (i) when solitons own different velocities, oblique interactions will occur; (ii) when solitons share the same velocity and keep a adjacent distance, they will attract and repel each other periodically. Namely, the bound states of solitons will emerge~\cite{PRE1997,Science1999}; (iii) when solitons propagate with the same velocity and keep a large distance, parallel solitons will happen. These three kinds of interactions will be illustrated respectively.

Assuming $k_{1}=k_{1R}+i\,k_{1I}$ and $k_{2}=k_{2R}+i\,k_{2I}$, we rewrite the soliton velocities as
\begin{eqnarray}
V_{1}=\frac{\rho\rho^{*}-8}{4[k_{1R}^{2}+(k_{1I}+1)^{2}]},\hspace{0.5cm}
V_{2}=\frac{\rho\rho^{*}-8}{4[k_{2R}^{2}+(k_{2I}+1)^{2}]},
\end{eqnarray}
while the subscripts $R$ and $I$ represent the real and imaginary parts, with $k_{1R}$, $k_{1I}$, $k_{2R}$ and $k_{2I}$ being real constants. The distances between solitons can be influenced by $|\Delta{\nu}|=|\nu_{1}-\nu_{2}|$.\\

\begin{center}
{\includegraphics[scale =
0.8]{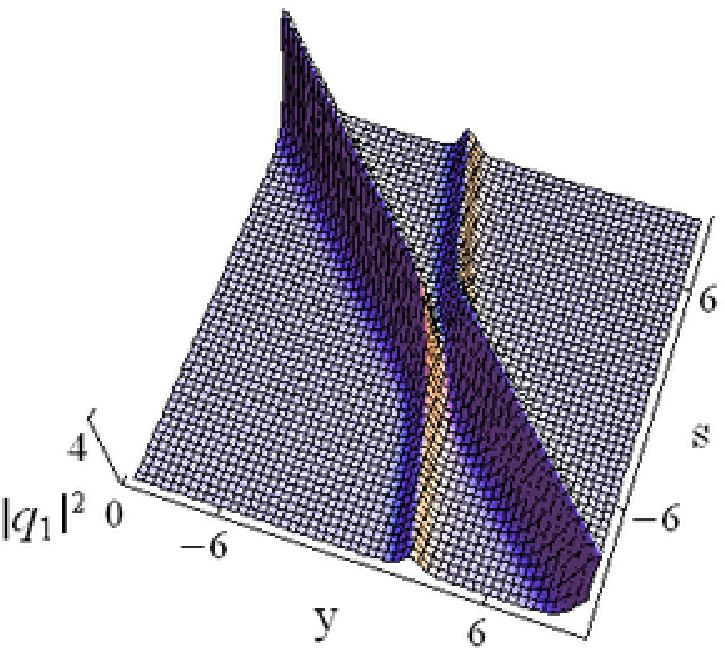}}\hspace{1.5cm}{\includegraphics[scale =
0.8]{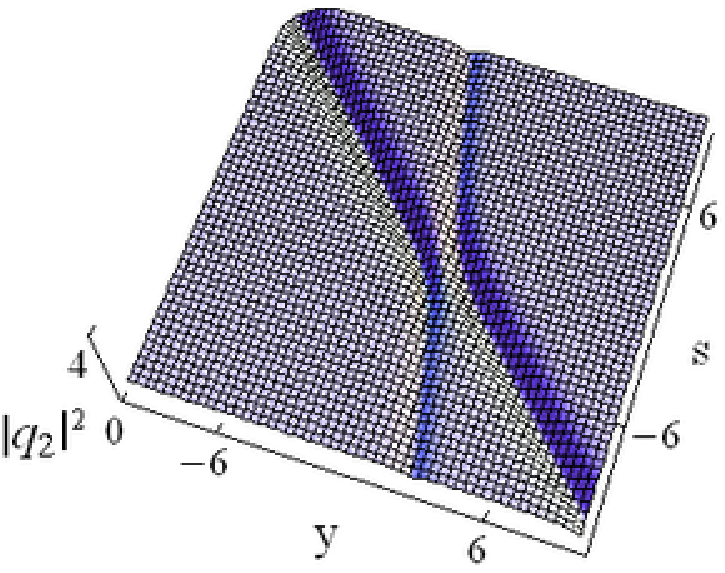}}
\ \\{\footnotesize\hspace{1cm}(a)  \hspace{6.5cm}(b) }
\ \\\flushleft{\footnotesize {\bf Figs.~1}
Oblique interactions between the two solitons via Solutions~(\ref{two-soliton}) with $k_{1}=1$, $k_{2}=2+i$, $\delta_{1}=2$, $\delta_{2}=2$, $\nu_{1}=2$, $\nu_{2}=0$ and $\rho=1$.}
\end{center}

\begin{center}
{\includegraphics[scale =
0.8]{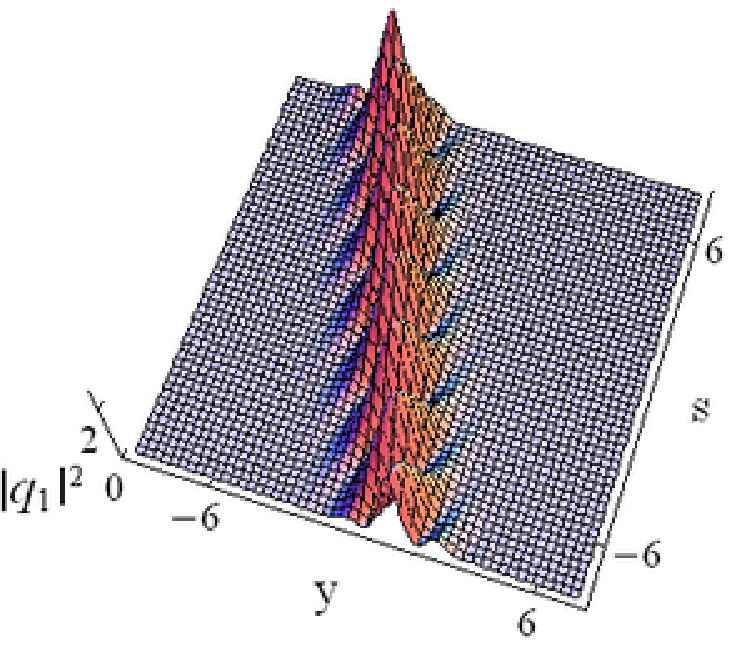}}\hspace{1.5cm}{\includegraphics[scale =
0.8]{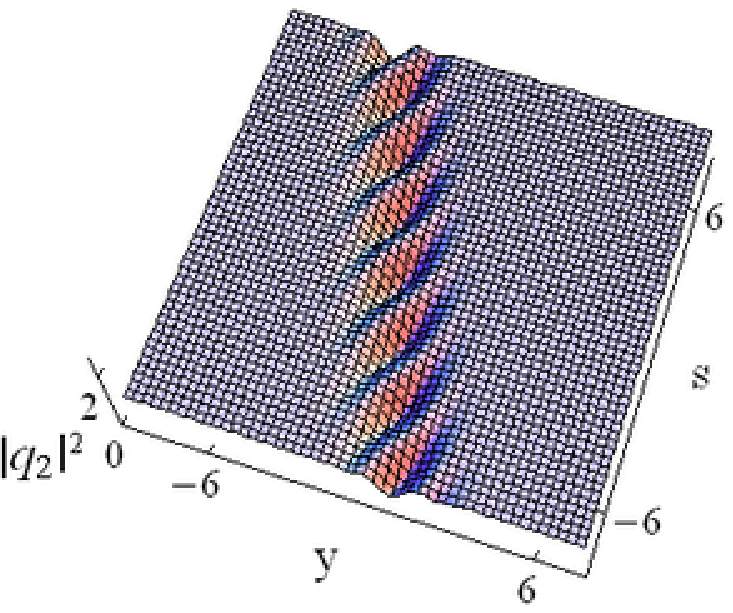}}
\ \\{\footnotesize\hspace{1cm}(a)  \hspace{6.5cm}(b) }
\ \\\flushleft{\footnotesize {\bf Figs.~2}
Bound states of solitons via Solutions~(\ref{two-soliton}) with $k_{1}=1+i$, $k_{2}=1.1-\frac{\sqrt{379}+10}{10}i$, $\delta_{1}=2$, $\delta_{2}=2$, $\nu_{1}=2$, $\nu_{2}=2$ and $\rho=1$.}
\end{center}
\begin{center}
{\includegraphics[scale =
0.8]{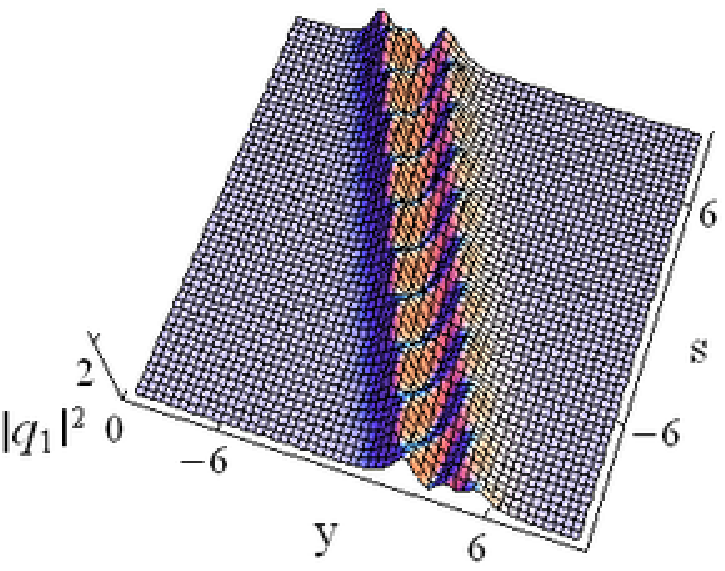}}\hspace{1.5cm}{\includegraphics[scale =
0.8]{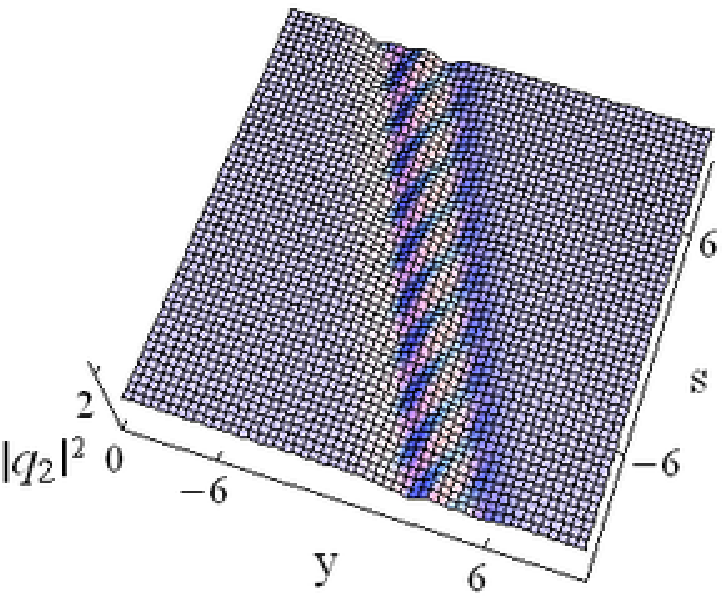}}
\ \\{\footnotesize\hspace{1cm}(a)  \hspace{6.5cm}(b) }
\ \\\flushleft{\footnotesize {\bf Figs.~3}
Bound states of solitons via Solutions~(\ref{two-soliton}) with the same parameters as those in Figs.~2 except for $\nu_{1}=-1$.}
\end{center}

\begin{center}
{\includegraphics[scale =
0.8]{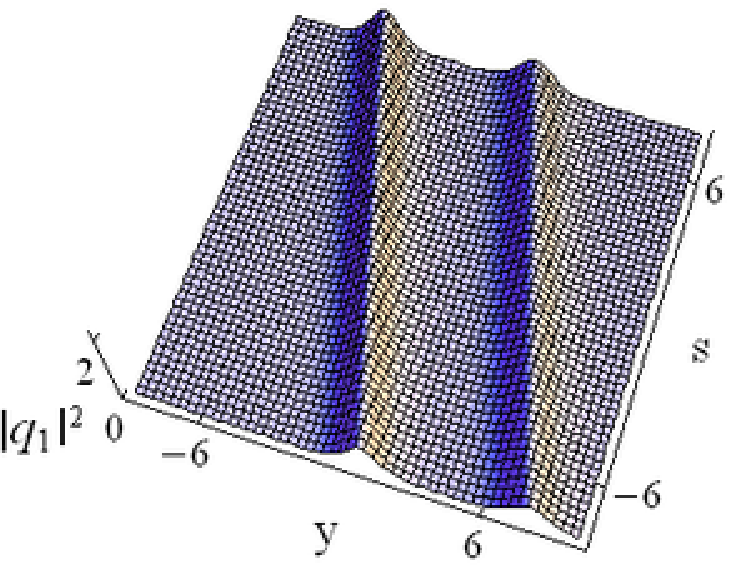}}\hspace{1.5cm}{\includegraphics[scale =
0.8]{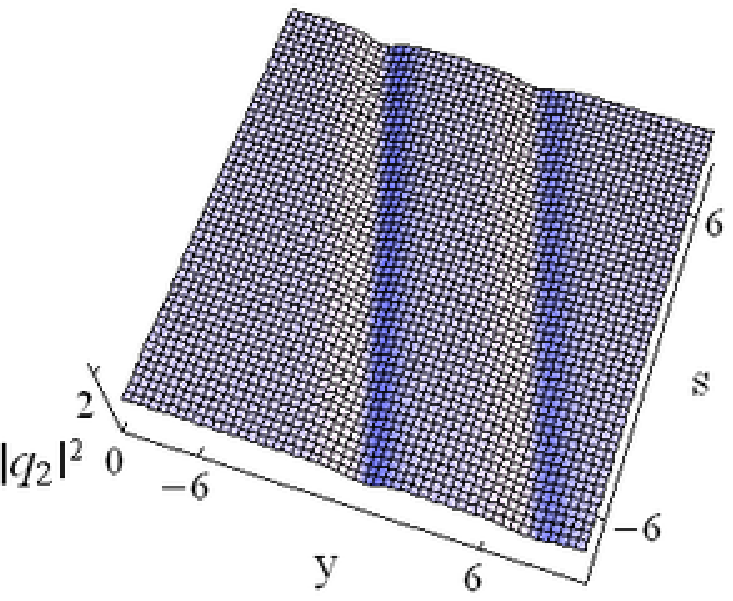}}
\ \\{\footnotesize\hspace{1cm}(a)  \hspace{6.5cm}(b) }
\ \\\flushleft{\footnotesize {\bf Figs.~4}
Parallel solitons via Solutions~(\ref{two-soliton}) with the same parameters as those in Figs.~2 except for $\nu_{1}=-5$.}
\end{center}

Figs.~1 displayed the oblique interactions between the two solitons which possess the velocities $V_{1}=-\frac{7}{8}$ and $V_{2}=-\frac{7}{32}$, respectively. After the interactions, solitons velocities and amplitudes keep invariant, while certain phase shifts occur. When we make the velocities as $V_{1}=V_{2}=-\frac{7}{20}$, the bound states of solitons and parallel solitons will happen, as seen as in Figs.~2-4. At the case of $|\Delta{\nu}|=0$, two solitons are bounded together that we cannot make a distinction between them, which can be seen in Figs.~2. When the solitons have the distance of $|\Delta{\nu}|=3$ in Figs.~3, interaction strength between the left soliton with large amplitude/depth and the right soltion with small amplitude/depth becomes weaker than in Figs.~2. When the parameters are chosen as $|\Delta{\nu}|=7$,  the interaction strength between solitons tends to be neglected. As seen in Figs.~4, two parallel solitons propagate mutual independently.\\

\noindent\textbf{5. Conclusions}\\

In this paper, with the help of Hirota method, we have investigated the coupled complex short pulse equations, i.e., Eqs.~(\ref{coupled-complex-sp}), which describe the propagation of ultra-short optical pulses in cubic nonlinear media. The following results should be mentioned:

(i) The bright-dark One- and Two-Soliton Solutions~(\ref{one-soliton}) and~(\ref{two-soliton}) have been explicitly expressed based on Bilinear Forms~(\ref{bilinear-forms}) with the Scale Transformations~(\ref{scale-transformation}).

(ii) Asymptotic Analysis~(\ref{before-interaction-1a})-(\ref{after-interaction-2b}) have been made on Solutions~(\ref{two-soliton}) to investigate the interactions between the two solitons. From Table~1, we conclude that interactions between two bright or two dark solitons are both elastic.

(iii) Oblique interactions~[see Figs.~1], bound states of solitons~[see Figs.~2 and 3] and parallel solitons~[see Figs.~4] have been illustrated analytically and graphically. Types of interactions can be influenced by soliton velocities and distances between solitons.

We hope that our study will be helpful to the research and development of optical fiber communications.\\

\noindent\textbf{Acknowledgments}\\

We express our sincere thanks to all members of the discussion team for their helpful suggestions.\\

\noindent\textbf{Appendix}\\

\noindent\text{Appendix A. Proof of Bilinear Forms~(\ref{bilinear-forms})}

Dividing the Bilinear Forms~(\ref{bilinear-forms}) by $f^{2}$, we get
\begin{subequations}\label{proof}
\begin{align}
&\left(\frac{g_{1}}{f}\right)_{ys}+\frac{g_{1}}{f}\cdot2(\ln\,f)_{ys}+i\left(\frac{g_{1}}{f}\right)_{s}
-i\left(\frac{g_{1}}{f}\right)_{y}=(\lambda+1)\frac{g_{1}}{f},\label{bilinear-form-proof-a}\\
&\left(\frac{g_{2}}{f}\right)_{ys}+\frac{g_{2}}{f}\cdot2(\ln\,f)_{ys}+i\left(\frac{g_{2}}{f}\right)_{s}
-i\left(\frac{g_{2}}{f}\right)_{y}=(\lambda+1)\frac{g_{2}}{f},\label{bilinear-form-proof-b}\\
&2(\ln\,f)_{ss}-2\lambda=\frac{1}{2}(|q_{1}|^{2}+|q_{2}|^{2}).
\end{align}
\end{subequations}
From the Scale Transformations~(\ref{scale-transformation}), we have
\begin{eqnarray}
\frac{\partial{x}}{\partial{s}}=2\lambda-2(\ln\,f)_{ss},\hspace{0.5cm}
\frac{\partial{x}}{\partial{y}}=\lambda-2(\ln\,f)_{sy},
\end{eqnarray}
i.e.,
\begin{eqnarray}
&&\partial_{y}=\varrho^{-1}\partial_{x},\hspace{0.5cm}\varrho^{-1}=\lambda-2(\ln\,f)_{sy},\notag\\
&&\partial_{s}=[2\lambda-2(\ln\,f)_{ss}]\partial_{x}-\partial_{t}=-\frac{1}{2}(|q_{1}|^{2}+|q_{2}|^{2})\partial_{x}-\partial_{t}.\notag
\end{eqnarray}
Eqs.~(\ref{coupled-complex-sp-a}) and~(\ref{coupled-complex-sp-b}) can be reduced from Eqs.~(\ref{bilinear-form-proof-a}) and~(\ref{bilinear-form-proof-b}), respectively.
We rewritten Eq.~(\ref{bilinear-form-proof-a}) as,
\begin{eqnarray}
\left[q_{1}\,e^{-i(y-s)}\right]_{ys}=\left[-2(\ln\,f)_{ys}+i\,\partial_{y}-i\,\partial_{s}+\lambda+1\right]q_{1}\,e^{-i(y-s)},
\end{eqnarray}
which means
\begin{eqnarray}
&&q_{1,ys}=\left[\lambda-2(\ln\,f)_{ys}\right]q_{1},\\
&&\varrho^{-1}\partial_{x}\left[-\partial_{t}-\frac{1}{2}(|q_{1}|^{2}+|q_{2}|^{2})\partial_{x}\right]q_{1}=\varrho^{-1}q_{1},\\
&&q_{1,xt}+\frac{1}{2}\left[(|q_{1}|^{2}+|q_{2}|^{2})q_{1,x}\right]_{x}+q_{1}=0.\label{coupled-complex-sp-a-proof}
\end{eqnarray}
Eq.~(\ref{coupled-complex-sp-a-proof}) is exactly equal to Eq.~(\ref{coupled-complex-sp-a}). In the similar way, Eq.~(\ref{coupled-complex-sp-b}) can be derived from Eq.~(\ref{bilinear-form-proof-b}).\\

\noindent\text{Appendix B. Relevant coefficients in Solutions~(\ref{two-soliton})}

The relevant coefficients in Solutions~(\ref{two-soliton}) are expressed as
\begin{eqnarray}
&&\lambda=-\frac{1}{4}\rho\,\rho^{*},\hspace{0.2cm}b=-1+\frac{1}{4}\rho\,\rho^{*},\hspace{0.2cm} \omega_{1}=\frac{4+4i\,k_{1}-\rho\,\rho^{*}}{4i+4\,k_{1}},\hspace{0.2cm}
\omega_{2}=\frac{4+4i\,k_{2}-\rho\,\rho^{*}}{4i+4\,k_{2}},\notag\\
&&\xi_{1}=\frac{k_{1}(-2i+k_{1}^{*})}{k_{1}^{*}(2i+k_{1})},\hspace{0.2cm}
\xi_{2}=\frac{k_{1}(-2i+k_{2}^{*})}{k_{2}^{*}(2i+k_{1})},\hspace{0.2cm}
\xi_{3}=\frac{k_{2}(-2i+k_{1}^{*})}{k_{1}^{*}(2i+k_{2})},\hspace{0.2cm}
\xi_{4}=\frac{k_{2}(-2i+k_{2}^{*})}{k_{2}^{*}(2i+k_{2})},\notag\\
&&\eta_{1}=-\frac{\delta_{1}\delta_{1}^{*}}{8\lambda+(\xi_{1}+\xi_{1}^{*})\rho\,\rho^{*}-4(\omega_{1}+\omega_{1}^{*})^{2}},\hspace{0.2cm}
\eta_{2}=-\frac{\delta_{1}\delta_{2}^{*}}{8\lambda+(\xi_{2}+\xi_{3}^{*})\rho\,\rho^{*}-4(\omega_{1}+\omega_{2}^{*})^{2}},\notag\\
&&\eta_{3}=-\frac{\delta_{2}\delta_{1}^{*}}{8\lambda+(\xi_{3}+\xi_{2}^{*})\rho\,\rho^{*}-4(\omega_{2}+\omega_{1}^{*})^{2}},\hspace{0.2cm}
\eta_{4}=-\frac{\delta_{2}\delta_{2}^{*}}{8\lambda+(\xi_{4}+\xi_{4}^{*})\rho\,\rho^{*}-4(\omega_{2}+\omega_{2}^{*})^{2}},\notag\\
&&\delta_{3}=-\frac{(k_{1}-k_{2})[\delta_{2}\eta_{1}(k_{1}+k_{1}^{*})(2i+k_{2}-k_{1}^{*})-
\delta_{1}\eta_{3}(k_{2}+k_{1}^{*})(2i+k_{1}-k_{1}^{*})]}{(k_{1}+k_{1}^{*})(k_{2}+k_{1}^{*})(2i+k_{1}+k_{2})},\notag\\
&&\delta_{4}=-\frac{(k_{1}-k_{2})[\delta_{2}\eta_{2}(k_{1}+k_{2}^{*})(2i+k_{2}-k_{2}^{*})-
\delta_{1}\eta_{4}(k_{2}+k_{2}^{*})(2i+k_{1}-k_{2}^{*})]}{(k_{1}+k_{2}^{*})(k_{2}+k_{2}^{*})(2i+k_{1}+k_{2})},\notag\\
&&\eta_{5}=\frac{(k_{1}^{*}-k_{2}^{*})[\delta_{3}\eta_{2}(k_{2}+k_{1}^{*})(2i+k_{2}-k_{2}^{*})-
\delta_{4}\eta_{1}(k_{2}+k_{2}^{*})(2i+k_{2}-k_{1}^{*})]}{\delta_{1}(k_{2}+k_{1}^{*})(k_{2}+k_{2}^{*})(-2i+k_{1}^{*}+k_{2}^{*})},
\hspace{0.2cm}\xi_{5}=\xi_{1}\xi_{4}.\notag
\end{eqnarray}

\end{document}